\documentclass[preprint,showkeys,preprintnumbers,amsmath,amssymb,superscriptaddress]{revtex4-1}
\usepackage{color}
\usepackage{braket}
\usepackage{graphicx}
\usepackage{dcolumn}
\usepackage{bm}

\newcommand{\beq}{\begin{equation}\displaystyle}
\newcommand{\eeq}{\end{equation}}
\newcommand{\bit}{\begin{itemize}}
\newcommand{\eit}{\end{itemize}}
\newcommand{\ben}{\begin{enumerate}}
\newcommand{\een}{\end{enumerate}}
\newcommand{\bc}{\begin{center}}
\newcommand{\ec}{\end{center}}

\begin{document}


\title{Pulse delay and group velocity dispersion measurement in V-type electromagnetically induced transparency of hot $^{85}Rb$ atom}
\author{Bankim Chandra Das}
\email{bankim.das@saha.ac.in}
\affiliation{Saha Institute of Nuclear Physics, HBNI, 1/AF, Bidhannagar,
Kolkata -- 700064, India.}
\author {Dipankar Bhattacharyya}
\affiliation{Department of Physics, Santipur College, Santipur, Nadia, West Bengal, 741404, India.}
\author{Arpita Das}
\affiliation{Saha Institute of Nuclear Physics, HBNI, 1/AF, Bidhannagar,
Kolkata -- 700064, India.}
\author{Satyajit Saha}
\affiliation{Saha Institute of Nuclear Physics, HBNI, 1/AF, Bidhannagar,
Kolkata -- 700064, India.}
\author{Shrabana Chakrabarti}
\affiliation{Saha Institute of Nuclear Physics, HBNI, 1/AF, Bidhannagar,
Kolkata -- 700064, India.}
\author{Sankar De}
\email{sankar.de@saha.ac.in}
\affiliation{Saha Institute of Nuclear Physics, HBNI, 1/AF, Bidhannagar,
Kolkata -- 700064, India.}

\date{\today}

\begin{abstract}
Pulse delay with the group velocity dispersion (GVD) characteristics was studied in the V-type electromagnetically induced transparency in the hyperfine levels of $^{85}Rb$ atoms with a closed system configuration. The phase coherency between the pump and the probe laser beams was maintained. We studied the pulse delay and the group velocity dispersion characteristics with the variation of the pump Rabi frequency taking temperature as a parameter. We observed a maximum of $268$ $ns$ pulse delay for $21.24 MHz$ pump Rabi frequency at $55^0C$ temperature of the Rb vapour cell. For a better understanding of the experimental results, we have derived an analytical solution for the delay characteristics considering the thermal averaging. The analytical solution was derived for a three level V-type system. The theoretical plots of the delay and the group velocity dispersion show the same characteristics as we observed in the experiment. This analytical approach can be further generalized for the higher level schemes to calculate different quantities such as susceptibility, group velocity delay or group velocity dispersion characteristics.
 \end{abstract}
%
\maketitle
\section{Introduction}

Interaction of two coherent fields with the atomic media had led to many interesting phenomena like electromagnetically induced transparency (EIT) \cite{harris}, coherent population trapping (CPT) \cite{Arimondo}, lasing without inversion (LWI) \cite{harris1} etc. When a weak probe beam interacts with the atomic medium in the presence of a strong pump beam, a quantum interference phenomenon can happen. This leads to EIT, rendering the medium transparent to the field over a small frequency range. In this EIT region, the medium has high normal dispersion which creates a control over the group velocity of light.\\
To study the absorptive and the dispersive properties of the medium a $\Lambda$-type system was always preferred in comparison to the V-type system. There were several studies on the subluminal light propagation in the $\Lambda$-type system \cite{kasapi95,schmidt96,Hau99,Wu10} whereas V-type system needs to be explored further.  
It was assumed that the V-type EIT was due to the Aulter Townes splitting but a recent study showed that quantum destructive interference was behind the occurrence of EIT in the V-type system \cite{Cui11}. V-type system has importance in the generation of superluminal light with cold atomic ensembles without gain \cite{Joshi05}. Beil et \textit{al}. showed that in $Pr^{3+}:Y_{2}SiO_{5}$ with the V-type system, storage time can be increased as compared to the $\Lambda$-type system \cite{Beil08}. Studies showed that the velocity selective effect in the V-type EIT in Zeeman levels with a re-pumper configuration can change the time delay \cite{Li07}.
Apart from what is mentioned above, there are not much studies done on the absorptive and the dispersive properties in the V-type EIT in hyperfine levels, both experimentally and theoretically. In our earlier study we have shown that EIT will be formed for a V-type system in the closed transition \cite{BankimJCP}.

In this article we have studied both experimentally and theoretically the dispersive property of $^{85}Rb$ using a pulsed probe and a continuous pump beam. We observed subluminal light propagation in the V-type system with closed system configuration. We also studied the group velocity dispersion characteristics and saturation intensity dependency with the variation of pump power taking temperature as a parameter. We observed a maximum of $268$ $ns$ pulse delay with $21.24 MHz$ pump Rabi frequency for $55^{0}C$ cell temperature. It was observed that if we increase the number density of the atoms by increasing the temperature, the delay will increase. The group velocity dispersion characteristics was found to be opposite to the pulse delay characteristics. We also derived an analytical solution to study the characteristics of the slow light propagation and group velocity dispersion characteristics taking account of the thermal averaging. The analytical solution was derived for a three level V-type system. We found out that these results were similar to the numerically solved results. This theoretical approach can be generalized in higher level schemes to further derive analytical solutions of susceptibility, delay or group velocity dispersion characteristics.

We maintained the phase coherence between the probe and the pump beams since the time delay is dependent on the phase coherence between the beams. The delay will be less if the phase coherency is not present \cite{Bae10}. We choose $^{85}Rb$ $F=3\rightarrow F'$ transition to form a closed V-type system.

\section{Experiment}

\begin{figure}[h]
\centering
\includegraphics[scale=.4]{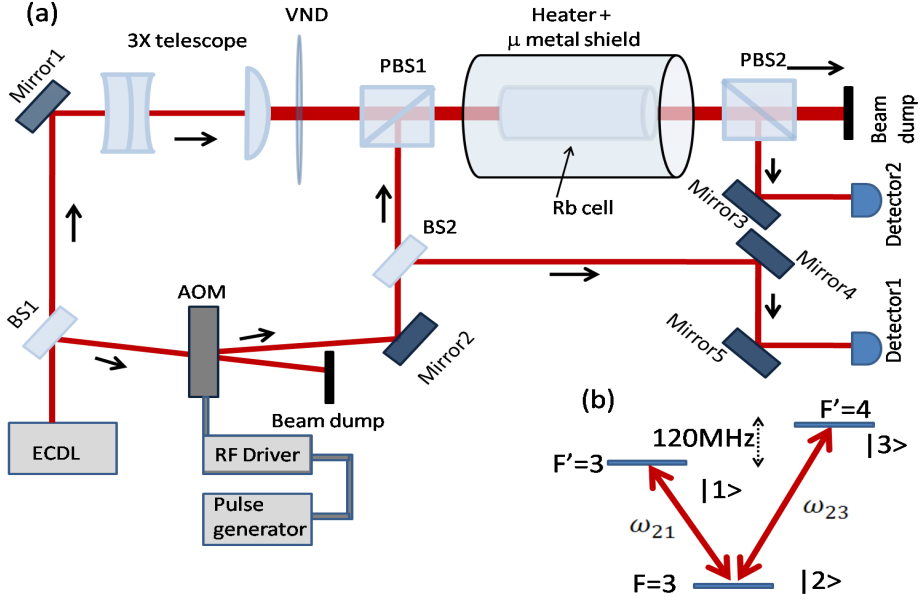}
\caption{(a) Experimental setup to observe slow light in a V-type electromagnetically induced transparency. ECDL: external cavity diode laser, PBS: polarizing beam splitter, BS: beam splitter, Rb cell: Rubidium cell, AOM: acousto optic modulator, VND: variable neutral density filter. (b) Energy level diagram for $^{85}Rb$ in $D_2$ transition to form a V-type system. $\omega_{21}$ and $\omega_{23}$ are the transition frequencies of the corresponding levels.}
\label{experimental setup}
\end{figure}

The experimental setup to study slow light in a V-type EIT is shown in figure \ref{experimental setup}(a). We have generated both the pump and the probe beams from a single external cavity diode laser (ECDL) to preserve the phase coherence between the beams. The ECDL was lasing at 780 nm (TOPTICA DL100). The output laser beam had a diameter of $\sim 2$ $mm$ and a line-width  $\sim 1$ $MHz$. A small part of the ECDL output was used for saturation absorption spectroscopy (SAS) (CoSy, TEM Messtechnik) (not shown in the figure \ref{experimental setup}(a)). SAS was further used for locking the laser frequency with the help of a lock-in-amplifier (LIR) and proportional integrator differentiator (PID) loop (TOPTICA LIR110). The remaining part of the laser beam was incident on the beam splitter (BS1). The reflected part from BS1 was used for the probe beam and the transmitted part was used for the pump beam. The pump beam was magnified by three times using a Galilean telescope so that the beam diameter becomes $\sim 6mm$. The probe beam was incident on an acousto optic modulator (AOM, INTRACTION). We took the $-1^{th}$ order of the diffraction in order to make the probe frequency less than the pump, so that an EIT condition can be produced for the V-type system \cite{fulton}. Now the $-1^{th}$ order of the AOM output was incident on BS2. The reflected part was taken as the reference signal which is detected by Detector1. The transmitted part from BS2 was mixed with the pump beam in a polarizing beam splitter (PBS1). We took orthogonal polarizations for the pump and the probe beams. Both the beams were sent through a Rb cell co-linearly. After the cell they were separated by another PBS2. The probe beam was detected by the photo-diode (Detector2).

The pump beam was locked to the closed transition $F=3 \rightarrow F'= 4$ of $^{85}Rb$ and the probe beam was down shifted by $-120$ $MHz$ with the AOM so that it can be locked to the transition $F=3 \rightarrow F'= 3$ (see figure \ref{experimental setup}(b)). In this way a V-type system was formed maintaining the phase coherence between the two laser beams. A 50 mm long and 25 mm diameter
cylindrical Rb cell was used in the experiment containing both $^{87} Rb$ and $^{85} Rb$ atoms in their natural
abundance and with no buffer gas. The pressure of the cell is of the order of $10^{-7}$ Torr. The Rb cell was put inside a $\mu$-metal shield. The cell was heated with a home made double walled heating jacket. Hot water was circulated through the cell jacket and the temperature of water was controlled using a home made temperature controller circuit. The temperature fluctuation was maintained within $\pm 0.5^0C$. We have avoided the resistive heating techniques in order to avoid unwanted magnetic field effects. To vary the pump power we have used variable neutral density (VND) filter. For our experiment, we have used the probe beam as a pulsed beam and the pump beam as a continuous beam. For generating the probe pulse an arbitrary function generator (Tektronix AFG3052C) was used to modulate the amplitude of the RF driver of the AOM. The probe pulse was chosen to be a Gaussian pulse of width $2.25$ $\mu s$. 

Throughout the experiment the probe Rabi frequency ($\Omega_c$) was fixed to $7.74 MHz$. We have started the experiment with the pump Rabi frequency as $8.64 MHz$ which is comparable to the probe Rabi frequency and increased the pump Rabi frequency gradually. 

\section{Experimental results and Discussion}
\begin{figure}[h]
\centering
\includegraphics[scale=.4]{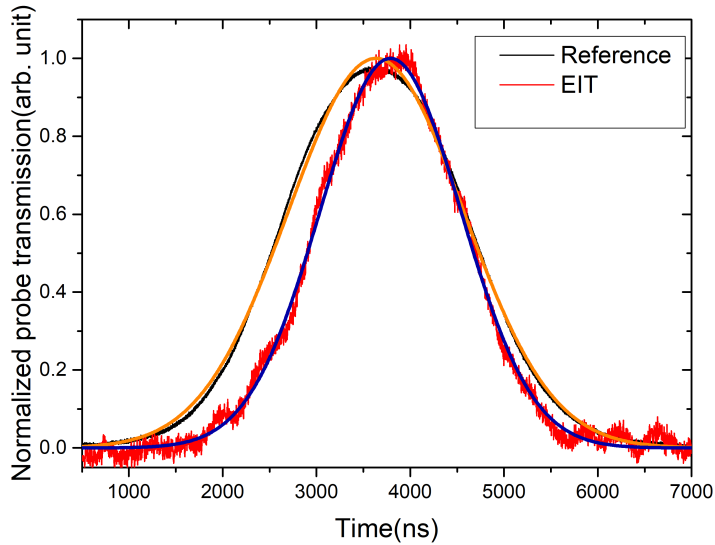}
\caption{Experimentally observed slow light pulse for pump Rabi frequency $34.68 MHz$ with $50^0C$. Observed delay is $154$ $ns$. Here the data are normalized in order to visualize the pulse delay and group velocity dispersion. The orange and the blue curves are the fitted curves of the reference and the EIT signals respectively.}
\label{spectrum}
\end{figure}

\begin{figure}[h]
\centering
\includegraphics[scale=.28]{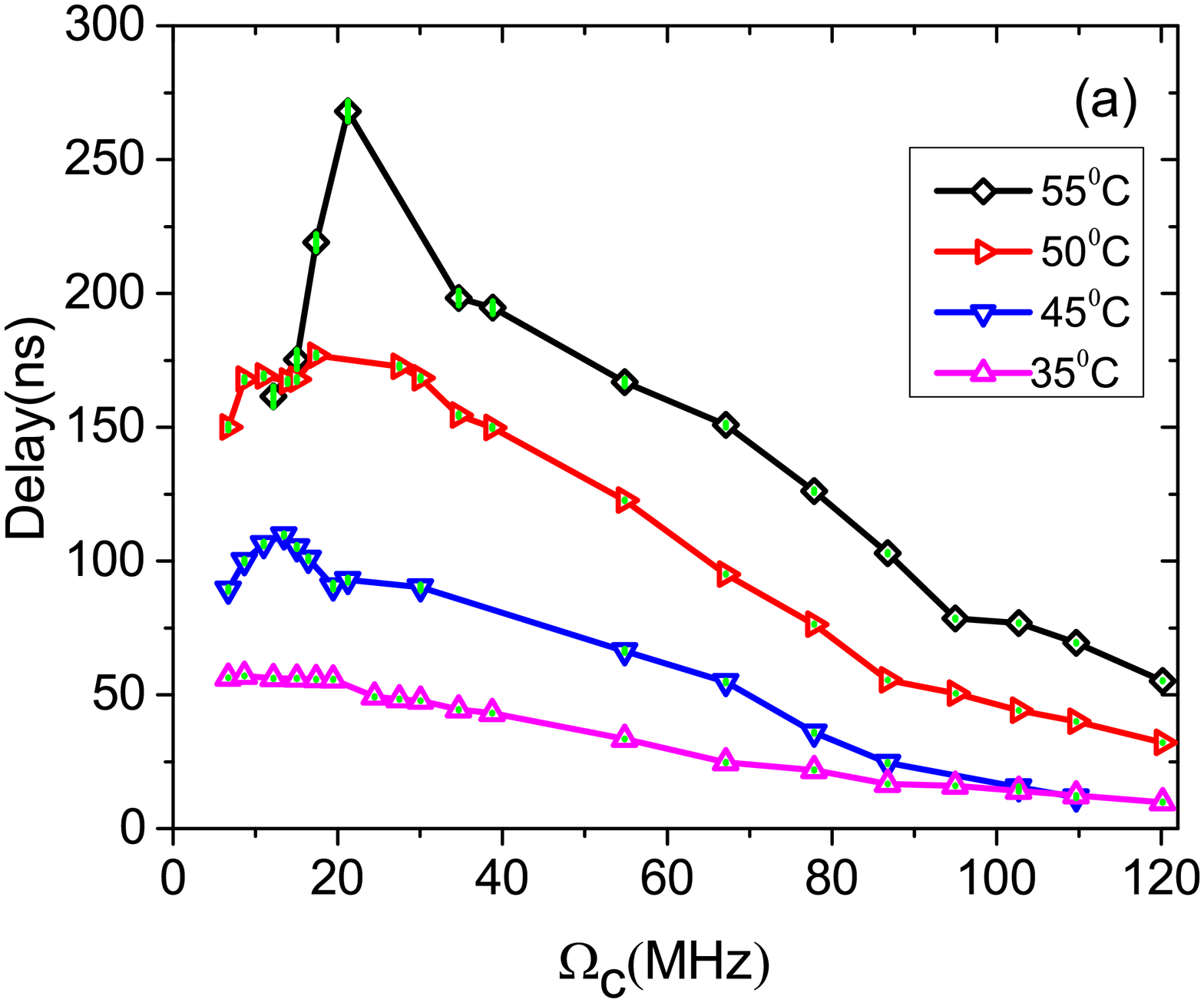}
\includegraphics[scale=.28]{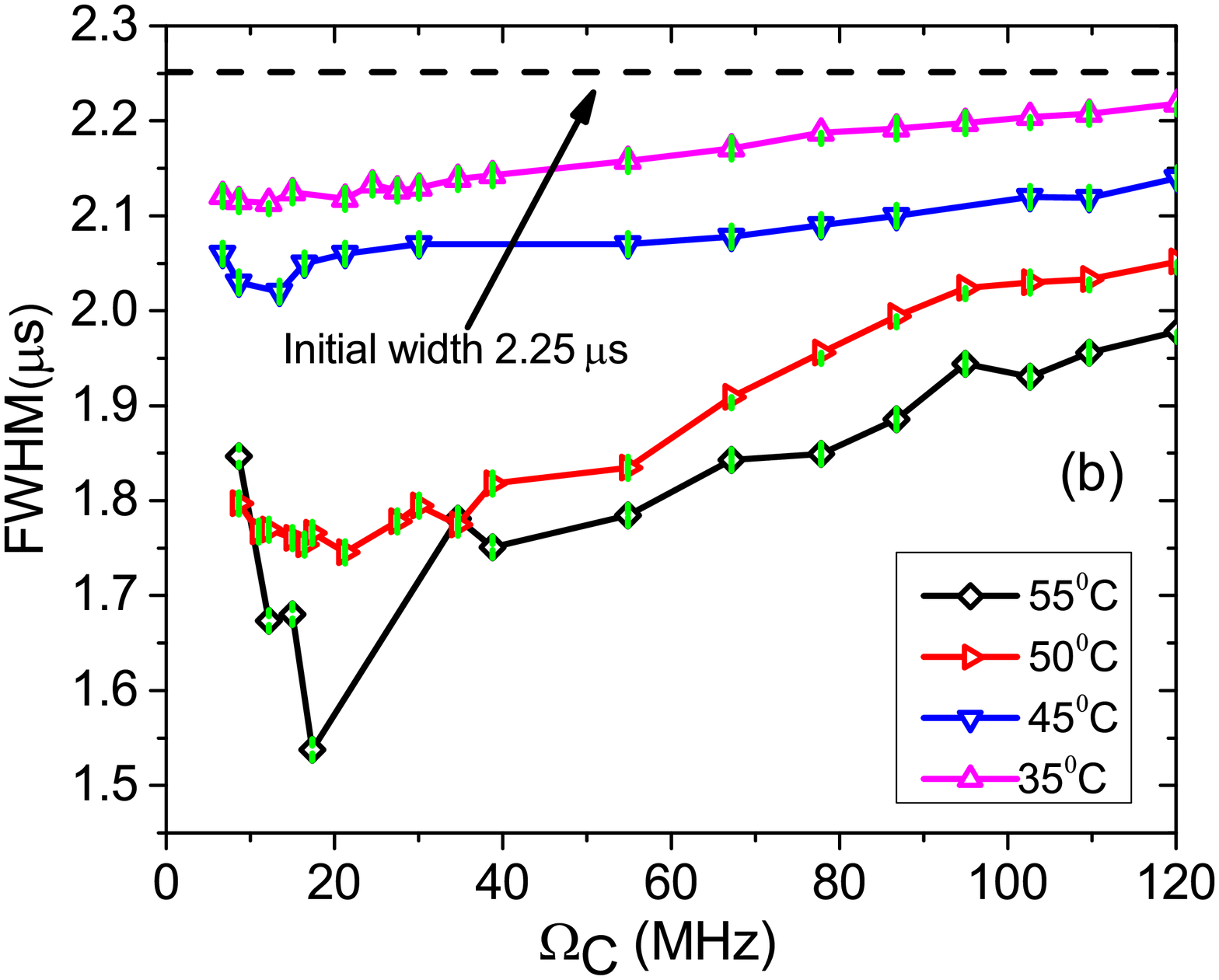}
\caption{(a) Variation of the pulse delay with $\Omega_c$ for different cell temperatures. (b) Variation of the FWHM of the output pulse with $\Omega_c$ for different cell temperatures. The lines are drawn for guidance of the eye.}
\label{delay and dispersion}
\end{figure}

We tuned the probe and the pump beams to get the EIT condition. Both the probe pulse, after passing through the cell and the reference pulse were detected by a pair of photo detectors simultaneously. The initial mismatch between the peaks of the pulses in both the  detectors was subtracted to get the correct delay values when the probe beam was in off resonant condition and the pump beam was switched off. 
 In figure \ref{spectrum}, the reference pulse and a probe  pulse output signal after the cell is shown for $34.68 MHz$ pump Rabi frequency and $50^0C$. The observed delay is $154$ $ns$. Here the data are normalized in order to visualize the delay and the FWHM of the pulse. We have taken an average of  $40$ shots so that the electronic and other noise fluctuations may be averaged out and we get a correct delay. The delay mentioned here is the peak separation between the two pulses which are fitted by a Gaussian function. We varied the pump Rabi frequency for a fixed temperature and observed the same characteristics for all the temperatures. We have measured the pulse delays for four different temperatures i.e., $35^0C$, $45^0C$, $50^0C$ and $55^0C$. Corresponding maximum delays were $57$ $ns$, $109$ $ns$, $176$ $ns$ and $268$ $ns$ respectively as shown in figure \ref{delay and dispersion}(a).

We observed that for a particular temperature, if we increase the pump Rabi frequency up to a certain value, the delay will start to increase but after reaching a maximum value, the delay will decrease if the pump Rabi frequency is increased further. It implied that if we increase the pump Rabi frequency till the saturation limit, the slope of the dispersion will increase and correspondingly the delay will increase. But after reaching the maximum delay, the pump power broadening starts to dominate resulting in a decreased slope of dispersion. Therefore the delay will be less if further pump Rabi frequency is increased. Theoretically we have derived an asymptotic solution for this phenomena (discussed in the theoretical section). When the pump Rabi frequency $\Omega_c \rightarrow 0$, we found that the pulse delay is increasing with a $\Omega_{c}^{2}$ dependency and when $\Omega_{c} \rightarrow \infty$, it decreases following a $\frac{1}{\Omega_{c}^{2}}$ relationship. In between these two limits, the delay is increasing almost linearly till it reaches the maximum value.

It was also observed that the pump Rabi frequency for which we got the maximum delay, was changing with the temperature. This can be explained on the basis of saturation limits. Since the number density increases with the temperature, the pump power needed to reach the saturation limit will also increase. So with the increase of temperature, the peak will shift. This can be seen in the delay versus pump Rabi frequency graph (figure \ref{delay and dispersion}(a)). We have shown later that our theoretical model also supports this statement. The maximum value of $\frac{d \chi}{d \omega}|_{\omega_0}$ is dependent on the optical density which is related to the temperature.\\

In figure \ref{delay and dispersion}(b), the group velocity dispersion i.e. the variation of the pulse width (FWHM) of the output pulse is shown with the pump Rabi frequency taking temperature as a parameter. It is noticed that the nature of the group velocity dispersion with the variation of pump Rabi frequency is just opposite to the delay characteristics. The pulse distortion or the group velocity dispersion can be understood in the following way. When a Gaussian pulse having an initial width $\sigma_{0}$ passes through a medium, its width is modified after interacting with the medium. From the response function formulation, it can be shown that (see the theory in Section 4), the effective width can be determined by, 
\begin{equation}
\sigma^2 = \sigma_{0}^2 + i k_{0}^{''} z
\end{equation}
Here, $k_{0}^{''}$ is the measure of group velocity dispersion which occurs due to the coherent effect of the EIT. Now for a normal dispersion $ k_{0}^{''} > 0$. But  $k_{0}^{''}$ is an imaginary quantity. Since in the EIT medium, the dispersion is normal, therefore $\sigma < \sigma_{0}$. It is expected that for higher delay, the pulse compression will be more. The experimentally observed FWHM curve of pulses shows the expected results. For higher temperature, the delay is higher resulting in a higher group velocity dispersion of the output pulse. For the maximum delay at a particular temperature the compression will be maximum which is also observed in figure \ref{delay and dispersion}(b).
Our theoretical simulation shows the same group velocity dispersion characteristics (see figure \ref{all delki with different temperature}(b)).\\
In our theoretical model we have considered only the two photon coherent effects in the GVD calculation. But in general the non negligible one photon Doppler absorption and effects of the higher order GVD terms (e.g. $\dfrac{d^3\chi}{d\omega^3}_{\omega=\omega_0}$ ) are also present in the experiment which we have ignored for simplicity of the calculation.
\section{Theoretical model}
To support our experimental observations we have simulated the output pulse envelope. To analyze the nature of the pulse delay and the group velocity dispersion we have calculated the response function of the atomic medium in the EIT condition.\\
Let us assume that the initial electric field of the probe is $E_{p}(0,t)$. Now we need to calculate the final electric field after it has traveled a distance $z$  through the atomic medium having a response function in frequency domain $H(z,\omega)$. 
From the response theory we can calculate the final output as,
\begin{equation}
E_{p}(z,\omega) = H(z,\omega)E_{p}(0,\omega)
\end{equation}
or in the time domain,
\begin{equation}
E_p(z,t) = \frac{1}{2 \pi}\int_{-\infty}^{\infty} e^{-i \omega t} H(z,\omega)E_p(0,\omega) d\omega
\end{equation}
In this experiment we have send the probe beam as a pulsed beam. So, the initial pulse can be written in the time domain as,
 \begin{equation}
 F(0,t) = \exp{\left[-\frac{t^2}{2\sigma_{0}^2}\right]}
 \end{equation}

Here $\sigma_0$ is the initial width. The inverse Fourier transformation of the initial pulse can be written as,
\begin{equation}
F(0,\omega) =\sqrt{2\pi \sigma_{0}^{2}} e^{-\sigma_{0}^{2}\omega^2/2} 
\end{equation}

\begin{figure}[h]
\begin{center}
\includegraphics[scale=.4]{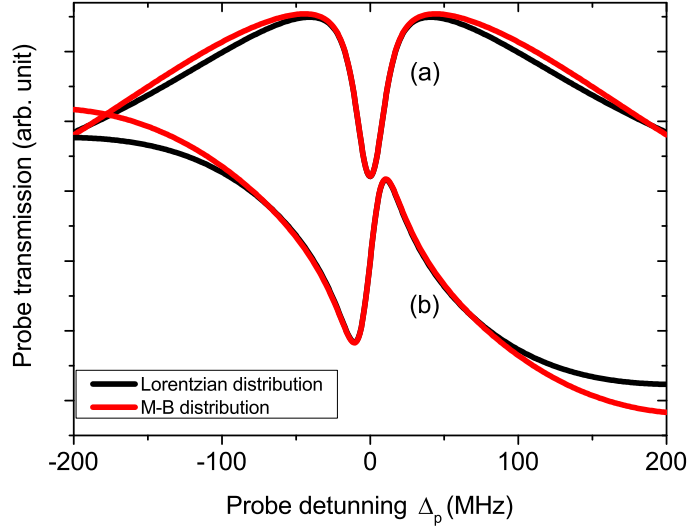}
\end{center}
\caption{Numerical simulations of the probe (a) absorption and (b) dispersion for two velocity distributions; M-B distribution (red) and Lorentzian distribution (black). Both the plots are plotted as a function of probe detunning ($\Delta_p$). Here, $\Omega_c= 10MHz$, $\Gamma= 6$ $MHz$.}
\label{MB_LZ_compare}
\end{figure}

Now the frequency response function of the envelope can be written as,
\begin{equation}
G(z,\omega) = e^{-i(k-k_0)z}
\end{equation}
Here $k$ is the wave vector. If we assume that the medium is behaving like a 2nd order  non-linear medium, we can Taylor expand $k(\omega)$ in the vicinity of the resonance frequency  $\omega_0$
\begin{equation}
k(\omega)= k_0 + k_{0}^{'} (\omega-\omega_0) + \frac{1}{2} k_{0}^{''}(\omega-\omega_0)^2 + ...
\end{equation} 
where, $k_{0}= k(\omega_0)$, $k_{0}^{'}=\frac{dk}{d\omega}|_{\omega_{0}}$ and $k_{0}^{''}= \frac{d^2k}{d\omega^2}|_{\omega_{0}}$. Furthermore, $k_{0}^{'}=\frac{dk}{d\omega}|_{\omega_{0}}= \frac{1}{v_g}$ where $v_g$ is the group velocity. This is the main reason behind the pulse delay in the medium and  $k_{0}^{''}$ is related to the pulse distortion or the group velocity dispersion \cite{Harris92}.
Now we can calculate $F(z,\omega)$ in the following way,
\begin{equation}
\begin{array}{ll}
F(z,\omega) &= G(z,\omega)F(0,\omega)\\

 &= \sqrt{2\pi \sigma_{0}^{2}} e^{-i k_{0}^{'} z \omega - i k_{0}^{''}\omega^2 z/2  -\sigma_{0}^{2}\omega^2/2}
\end{array}
\end{equation}
 Using the Fourier transformation, $F(z,t)$ becomes,
\begin{equation}
F(z,t)= \sqrt{\frac{\sigma_{0}^{2}}{\sigma_{0}^{2} + i k_{0}^{''}z}}\exp\left[-\dfrac{(t-k_{0}^{'}z)^2}{2(\sigma^2_{0} +  i k_{0}^{''}z)}\right]
\end{equation}
This output pulse is detected by the detector. The above expression shows that the pulse will be  delayed by a time $k_{0}^{'}z$ if it has traveled a distance z in a medium. The effective time delay  $\delta t$ can thus be calculated by subtracting the time required to travel a distance z in the vacuum as given by $\delta t = \frac{z}{v_g}- \frac{z}{c} \simeq \frac{z}{v_g}$. The width is also modified after passing through the medium. The effective width can be determined by, 
\begin{equation}
\sigma^2 = \sigma_{0}^2 + i k_{0}^{''} z
\end{equation}
So the group velocity dispersion is arising due to the $k_{0}^{''}$ term. The wave vector $k(\omega)$ is related to the refractive index i.e., $k(\omega)= \frac{n \omega}{c}$ or more generally to the susceptibility of the medium, which can be calculated by the density matrix model. Now $k^{'} = (1 + \frac{\omega_0}{2}\frac{d\chi}{d\omega}|_{\omega_o}
)/c$ and $k{''}= \frac{\omega_{0}}{2c}\frac{d^{2}\chi}{d\omega^2}|_{\omega_{0}}$. Since in our experiment we are considering a three level V-type system as shown in figure \ref{experimental setup}(b), we have solved the master equation for a three level system. The master equation is given by \cite{scully},

\begin{equation}\label{master eq}
\frac{d}{dt}\rho = \frac{-i}{\hbar} [H, \rho]
\end{equation}
Here $H$ is the total Hamiltonian of the system and $\rho$ is the density operator. These $H$ and $\rho$ are $3\times3$ matrices. We have added the decay terms phenomenologically. Since our pump beam was continuous we have solved the density matrix equation in steady state condition. The probe pulse can be calculated using the coherence term i.e., $\rho_{21}$. The coherence term $\rho_{21}$ can be solved in steady state condition to get the susceptibility of the medium. $\rho_{21}$ can be written as,

\begin{equation}
\rho_{21} = \frac{i \Omega_p}{2} \dfrac{\left(\Gamma^{2} + 4(\Delta_{c} + kv)^2 + 2 \Omega_{c}^{2}\right)-\Omega_{c}^{2} \dfrac{\Gamma^2/4 + (\Delta_{c} + kv)^2}{\left[\Gamma/2+ i (\Delta_c + kv)\right]\left[\Gamma- i (\Delta_p -\Delta_c)\right]}} {(\Gamma^2+ 4(\Delta_{c} + kv)^2+ 4\Omega_{c}^{2})\left[\Gamma/2 -i (\Delta_p + k v)+ \dfrac{\Omega^2_{c}/4 }{\Gamma - i(\Delta_p- \Delta_c)}\right]}
   \end{equation}
Here $\Omega_{p}$, $\Omega_{c}$, $\Gamma$ are the probe Rabi frequency, the pump Rabi frequency and the natural line-width respectively. $\Delta_p$ is the  probe detuning which is defined as $\omega_{21}-\omega$. $\omega_{21}$ is the resonant frequency of the transition $\ket{2}\rightarrow \ket{1}$  and $\omega$ is the field frequency. $\Delta_c= \omega_{23}-\omega$ is the pump detunning. $\omega_{23}$ is the resonant frequency of the transition $\ket{2}\rightarrow \ket{3}$. The wave vector of the pump and the probe beams are almost equal.
In a vapor cell we need to take care of the velocity of the atoms to get the total absorption or dispersion. The atoms obey Maxwell-Boltzmann (M-B) velocity distribution. Considering all the possible velocities the susceptibility becomes,
\begin{equation}\label{chi}
\chi  = \dfrac{\mu}{\epsilon_{0} E_p}\int_{-\infty}^{\infty} N(k v) \rho_{21} d(k v)
\end{equation}
 The distribution for a velocity $v$ can be written as,
\begin{equation}
N(kv) = \frac{N_0}{\sqrt{\pi k^2 u^2}} e^{-(k v)^2/(ku)^2}
\end{equation}
 Here $u$ is the most probable velocity which is related to the Doppler width of the absorption. Equation \ref{chi} is not analytically solvable for the M-B velocity distribution. But instead of M-B velocity distribution which is a Gaussian distribution, if we assume the distribution to be a Lorentzian having the same FWHM as the Doppler i.e. $2 W_{D} =2 \sqrt{\ln{2}} ku$, then the above equation \ref{chi} can be solved analytically. The assumed Lorentzian velocity distribution \cite{javan02} is,
\begin{equation}
N (k v) = N_{0} \Lambda_0 \dfrac{W_{D}/\pi}{W_{D}^2 + (k v)^2}
\end{equation}
Here $N_0$ is the number density of the atoms, $k$ is the wave vector and $\Lambda_0$ is a constant. We observed that by taking $\Lambda_0$ to be $\sqrt{\pi \ln 2}$, if we solve equation \ref{chi} numerically assuming both the above mentioned distributions, the results are completely overlapped for both the absorption and the dispersion in the EIT region (figure \ref{MB_LZ_compare}). There is mismatch however on the outer side of the EIT i.e. the Doppler wing part of the simulated plots as shown in figure \ref{MB_LZ_compare}. Therefore we showed that we can use the Lorentzian velocity distribution in order to calculate $\chi$ and also to estimate the time delay and the group velocity dispersion characteristics analytically. \\

The integral in equation \ref{chi} can be solved by contour integral method, considering the Lorentzian distribution. $\chi$ has five poles, $\pm i W_{D} $, and $\pm i \sqrt{\Gamma^2/4 + \Omega_c^2}$ and $-(i \Gamma/2 +\Delta_{p} +\frac{i\Omega_{c}^{2}/4}{\Gamma-i \Delta_p} )$.  We shall consider the contribution of two poles in the upper half plane $+i W_{D} $, and $+ i \sqrt{\Gamma^2/4 + \Omega_c^2}$. Here we assumed that the pump is on resonance i.e. $\Delta_{c}=0$. Let us also assume that $\chi= \chi_{1} + \chi_{2}$ where $\chi_{1}$ is the contribution of the pole $+i W_{D} $ and $\chi_{2}$ is the contribution of the pole $+ i \sqrt{\Gamma^2/4 + \Omega_c^2}$. Since we are interested in $\dfrac{d\chi}{d \omega} =\dfrac{d\chi_{1}}{d \omega}+\dfrac{d\chi_{2}}{d \omega}$, the analytical expression for $\dfrac{d\chi}{d \omega}|_{\omega_{0}}$ is shown below. Here we have assumed $\omega_{21}=\omega_{0}$ for simplicity.

\begin{figure}[h]
\centering
\includegraphics[scale=.25]{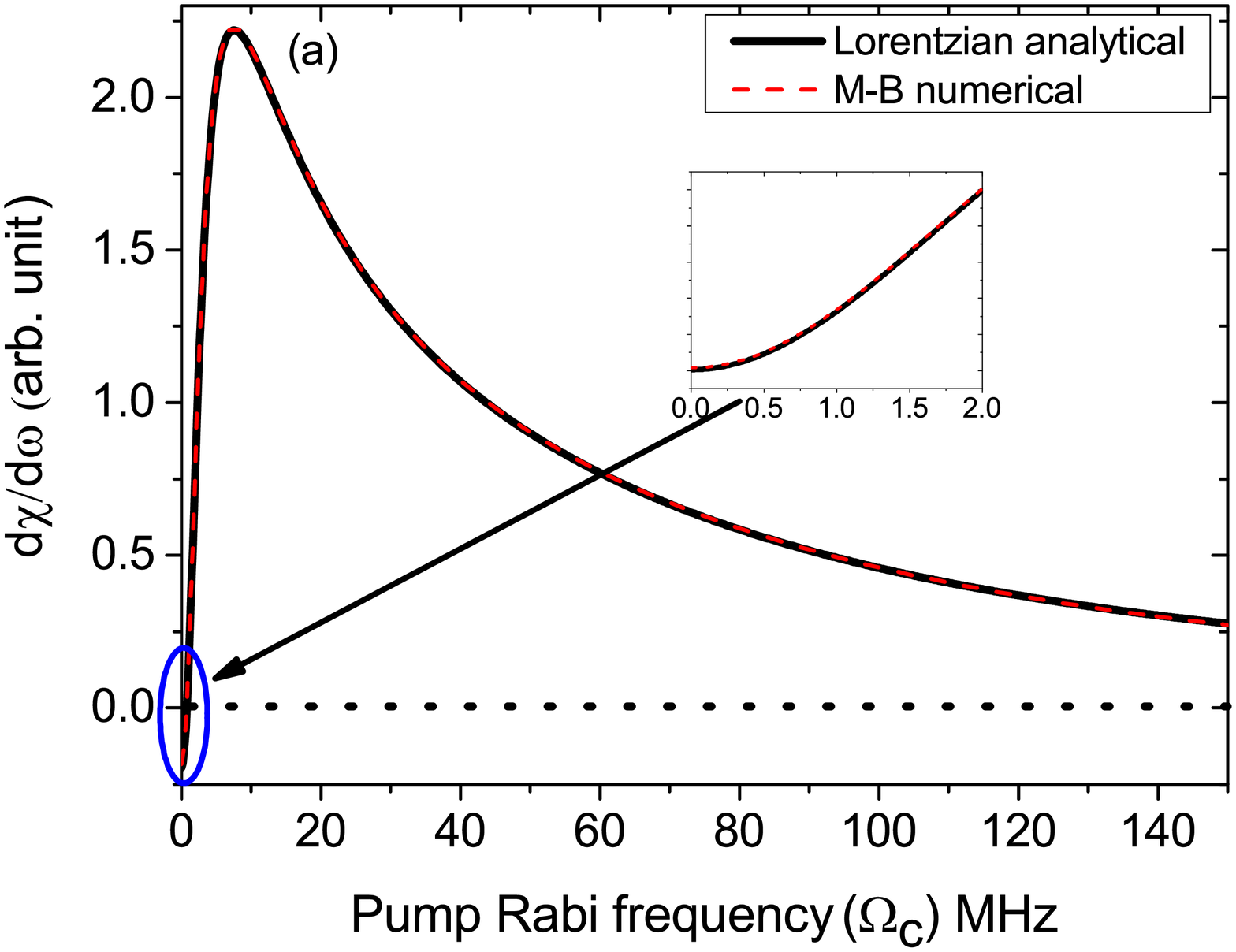}
\includegraphics[scale=.25]{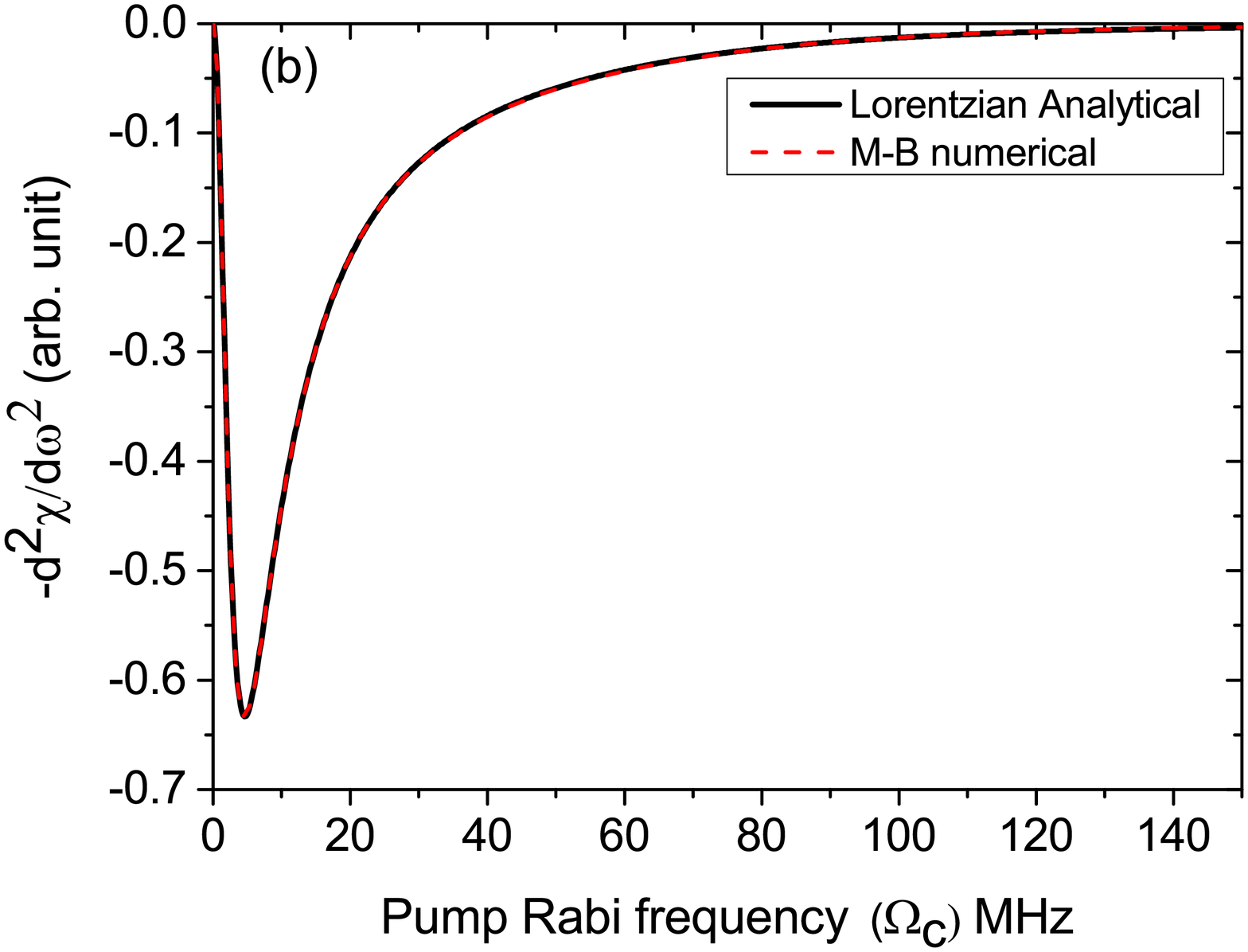}
\caption{Simulated plots for (a) $\frac{d\chi}{d\omega}|_{\omega_{0}}$ vs pump  Rabi frequency $\Omega_c$ and (b) $-\frac{d^2\chi}{d\omega^2}|_{\omega_{0}}$ vs pump Rabi frequency $\Omega_c$ using both the analytical solution considering Lorentzian distribution and the numerical solution considering M-B distribution.}
\label{all delki}
\end{figure}

The contribution of the pole $kv = + i W_{D}$ in $\frac{d\chi}{d\omega}$ when the pump and the probe beams are on resonance, is given by,
\begin{equation}\label{analytical delki1}
\dfrac{d\chi_{1}}{d\omega}|_{\omega_{0}}=-\dfrac{\mu\Omega_{p}}{\epsilon_{0} E_p}\times\dfrac{N_{0}\Lambda_{0}}{4W_{D}^{2}- \Gamma^2- 4\Omega_{c}^{2}} \times \dfrac{4(8 W_{D}^{2} \Gamma^{2}- 2\Gamma^{4} + 4 W_{D} \Gamma \Omega_{c}^2 - 2\Gamma^2 \Omega_{c}^2+ \Omega_{c}^{4})}{(2\Gamma^2 + 4W_{D} \Gamma + \Omega_{c}^2)^2}
\end{equation}
The contribution of the pole $kv = i \sqrt{\frac{\Gamma^2}{4}+ \Omega_{c}^2}$ in $\frac{d\chi}{d\omega}$ when the pump and the probe beam are on resonance, is given by, 
\begin{equation}\label{analytical delki2}
\dfrac{d\chi_{2}}{d\omega}|_{\omega_{0}}=-\dfrac{\mu\Omega_{p}}{\epsilon_{0} E_p}\times\dfrac{N_{0}\Lambda_{0}}{-4W_{D}^{2}+ \Gamma^2+ 4\Omega_{c}^{2}}\times \dfrac{8W_{D} \Omega_{c}^2 \left(6\Gamma^2 + \Omega_{c}^{2} + 2\Gamma \sqrt{\Gamma^2 + 4 \Omega^{2}_{c}}\right)}{\sqrt{\Gamma^2 + 4 \Omega^{2}_{c}}\left[2\Gamma^2 + \Omega_{c}^2 + 2\Gamma\sqrt{\Gamma^2 + 4 \Omega^{2}_{c}}\right]^2}
\end{equation}
The above equations \ref{analytical delki1} and \ref{analytical delki2} are the analytical solutions for the time delay characteristics. In figure \ref{all delki}(a), we have plotted the delay characteristics obtained from both the analytical solution following the Lorentzian distribution along with the numerically solved M-B distribution. Analytically obtained solution is completely overlapped with the numerical solution.

Now using the analytical solution the asymptotic behavior can be further solved. When $\Omega_{c} \rightarrow 0$, $\dfrac{d\chi}{d\omega}$  becomes,
\begin{equation}
\dfrac{d\chi}{d\omega}|_{\omega_{0}} \rightarrow -\dfrac{2\Lambda_{0}}{(2W_D + \Gamma)^2} +4\left(\dfrac{W_{D}\Lambda_{0}}{\Gamma^3(4W_{D}^2 -\Gamma^2)}- \dfrac{2\Lambda_{0}}{(2W_{D}-\Gamma) (2W_{D}+\Gamma)^3} \right) \Omega_{c}^2
\end{equation}

It can be observed that when $\Omega_c \sim 0 $, $\dfrac{d\chi}{d\omega}|_{\omega_{0}}$ increases as $\Omega_{c}^{2}$ (see figure 5(a)). But there is also a term $-\dfrac{2\Lambda_{0}}{(2W_D + \Gamma)^2}$, where the negative sign signifies that when $\Omega_c = 0$ it will reduce to one photon transition. In this case, the EIT will disappear giving a simple Doppler absorption profile. The slope of the dispersion will become negative here. Experimentally we have worked only in the EIT regime. So the lower pump Rabi frequency part is not observed in the experimental curves.\\
When $\Omega_c \rightarrow \infty $ it falls down as,
 \begin{equation}
 \dfrac{d\chi}{d\omega}|_{\omega_{0}} \rightarrow  \dfrac{\Lambda_{0}}{ \Omega_{c}^{2}}
 \end{equation}
 Overall the delay characteristics can be summarized as; when $\Omega_c \sim 0 $, it increases as $\Omega_{c}^{2}$ and then it linearly increases till the saturation occurs where it becomes maximum. After that when pump broadening starts dominating, the slope of the dispersion curve starts to decrease with $\frac{1}{\Omega_{c}^{2}}$. These theoretically simulated graphs support our experimental observations quite well. 
 It can be seen from the analytical expression (equation \ref{analytical delki1}  and \ref{analytical delki2}) that the maximum delay is dependent on the most probable velocity $u$, which is further related to the temperature of the system. So, if we increase the temperature of the cell, the pump Rabi frequency for which the maximum delay observed will change. This is also observed in the experiment.\\
From the analytical solution of $\chi$ we can calculate $\dfrac{d^2\chi}{d \omega^2}|_{\omega_0}$ to get the group velocity dispersion.
In figure \ref{all delki}(b) we have shown both the numerically calculated values with the M-B distribution and the analytical solution with the Lorentzian distribution. This is showing the same feature as that of the experimentally observed spectrum. As we mentioned earlier that for the highest delay the group velocity dispersion will be the maximum. Here also the analytically obtained results completely overlapped with the numerically solved results. The $\dfrac{d^2\chi}{d \omega^2}$ is itself imaginary and the modified width of the pulse $\sigma^2 = \sigma_{0}^2 + i k_{0}^{''} z$ carries a imaginary $i$ before $k_{0}^{''}$. To avoid the confusion with the experimental plot (figure \ref{delay and dispersion}(b)) we have plotted $-\dfrac{d^2\chi}{d \omega^2}|_{\omega_0}$ vs pump Rabi frequency $(\Omega_c)$ in figure \ref{all delki}(b).

\begin{figure}[h]
\centering
\includegraphics[scale=.25]{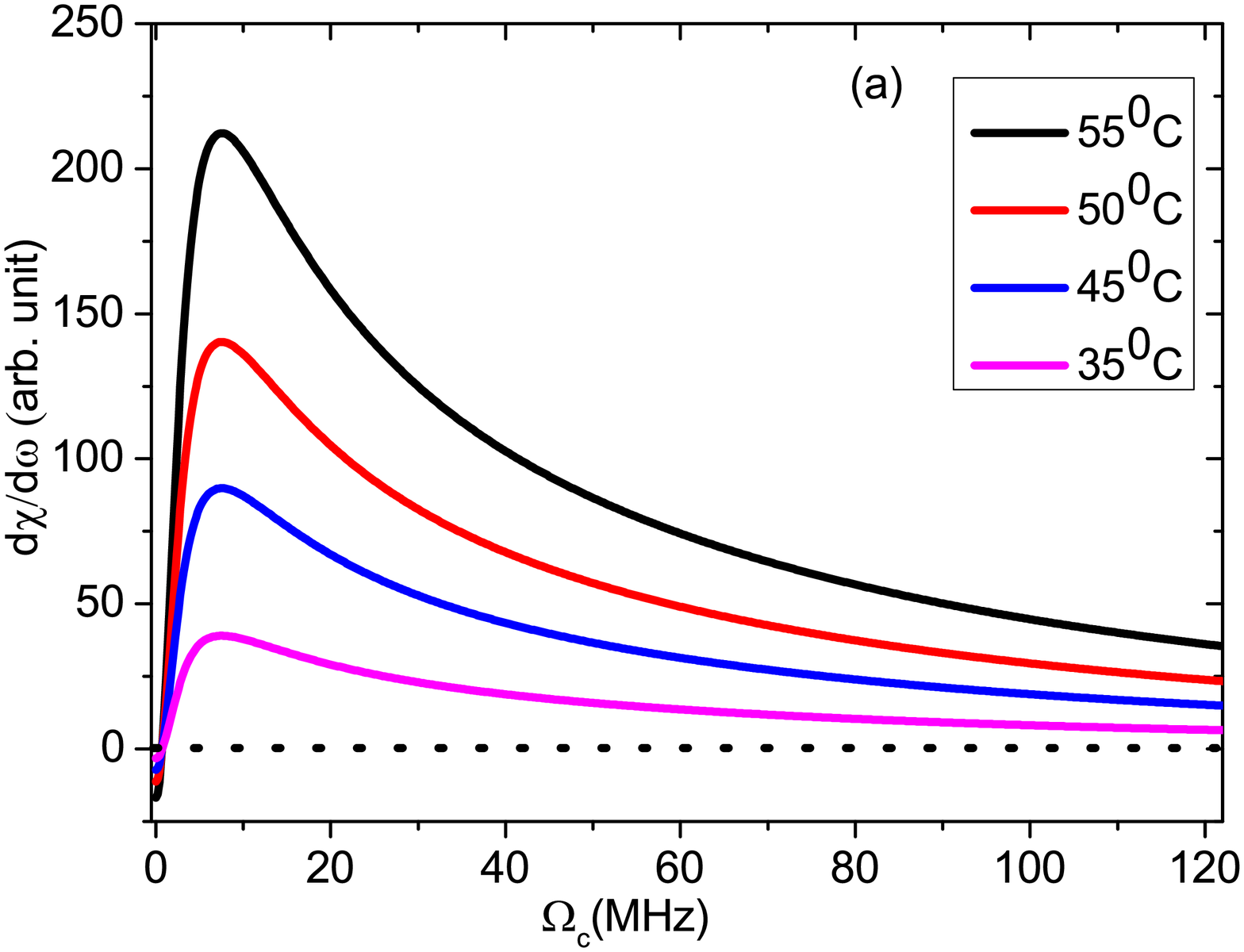}
\includegraphics[scale=.25]{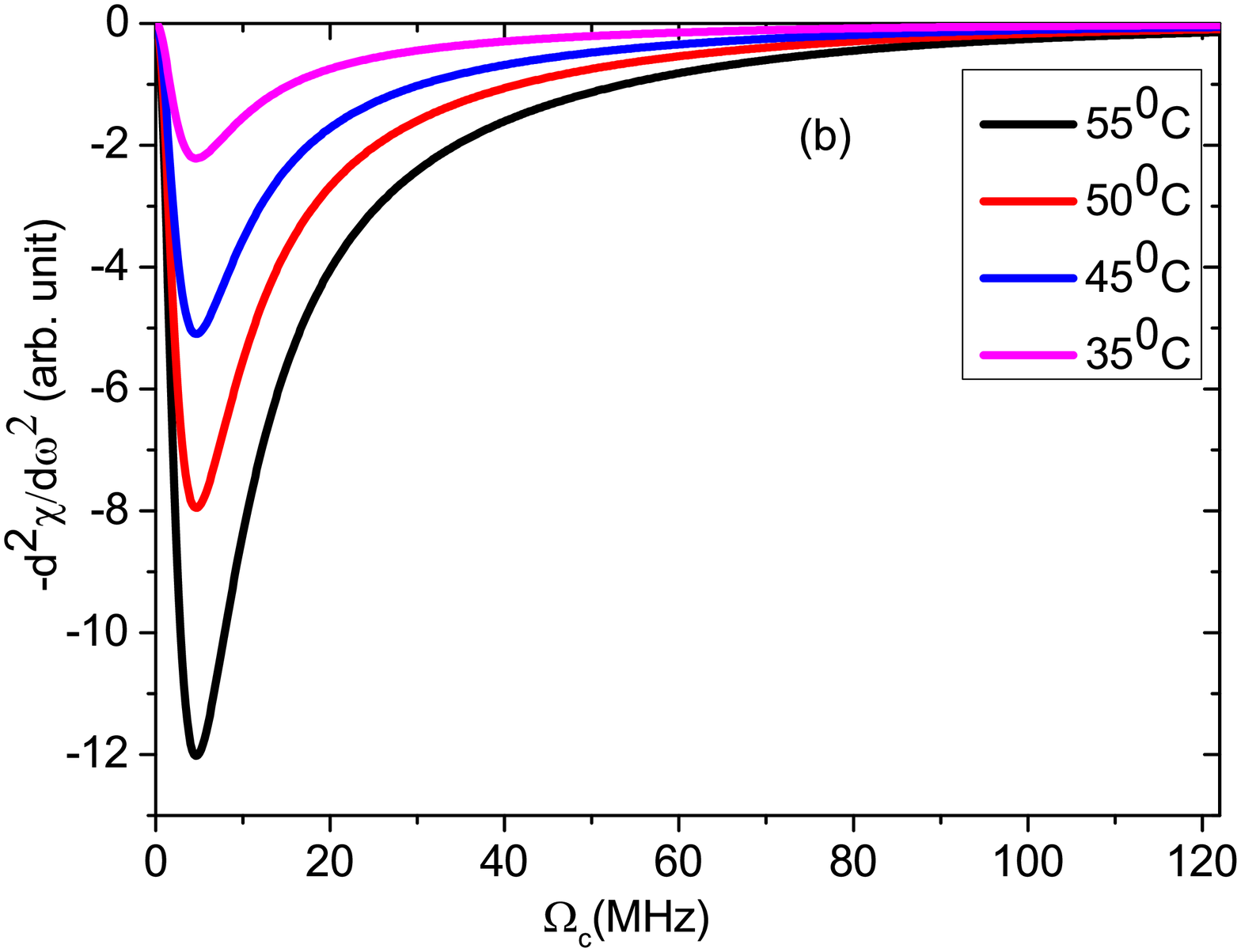}
\caption{Simulated plots for (a) $\frac{d\chi}{d\omega}|_{\omega_{0}}$ vs $\Omega_c$ and (b) $-\frac{d^2\chi}{d\omega^2}|_{\omega_{0}}$ vs $\Omega_c$ using the analytical solution considering Lorentzian distribution taking temperature as parameter (see legends).}
\label{all delki with different temperature}
\end{figure}

 In order to compare the experimental features with the theoretical ones, we have calculated the $\frac{d\chi}{d\omega}|_{\omega_{0}}$ vs  $\Omega_c $ and $-\frac{d^2\chi}{d\omega^2}|_{\omega_{0}}$ vs $\Omega_c $ for different temperatures considering our analytical results of $\chi$ as shown in figures \ref{all delki with different temperature} (a and b). These are giving similar characteristics as observed in the experiment. During the experiment when we had heated up the cell, a small portion of the $Rb$ vapor had deposited on the cell window. For this reason we were unable to calculate the atom number density $N_0$ exactly. Rather we took standard values of $N_0$ for a specific temperature.
\section{Conclusion}
We experimentally observed subluminal light propagation in a V-type system under the EIT condition. Pulse delay and group velocity dispersion characteristics of the input pulse were studied with the variation of the pump Rabi frequency taking temperature as a parameter.
It is observed that pump Rabi frequency for which the maximum pulse delay is observed, is a function of temperature.
 Group velocity dispersion is studied with the variation of the pump Rabi frequency. To support our experimental observations, we have formulated an analytical expression for the susceptibility considering thermal averaging. The simulated spectra shows the same nature as that of the experimentally observed results.

Although we have done the experiment in the closed transition, the Doppler effect of the hot atomic vapour would induce the non-closed transitions $F=3\rightarrow F'=2, 3$ and pump population from the closed transition $F=3\rightarrow F'=4$. Due to this optical pumping effect, the population will be reduced for the closed transition. Therefore the EIT effect will be diminished. This results in a further reduction of the slowing effect.


This pulse delay and group velocity dispersion study in the V-type system have several importance in the application of storage of information. 
 V-type system can be used for the quantum beat generation which is not possible in a $\Lambda$-type system \cite{Yamamoto98}. 
Moreover our theoretical approach can be generalized further in higher level systems for an analytical solution.

\section{Funding}
Department of Atomic Energy, Govt. of India (DAE) (Grant No. 12-R\&D-SIN-5.02-0102); Department of Science and Technology (DST), Govt. of India (Sanction no. SR/WOS-A/PM-1040/2014(G)).

\end{document}